\documentclass[9pt]{article}
\usepackage{spconf,amsmath,graphicx}
\usepackage{multirow}
\usepackage{hyperref}
\usepackage{booktabs}
\usepackage{amssymb}
\usepackage{caption}
\usepackage{float}
\usepackage{lipsum}

\title{ENRICHING UNDER-REPRESENTED NAMED-ENTITIES TO IMPROVE SPEECH RECOGNITION PERFORMANCE}
\name{%
\parbox{0.9\linewidth}{\centering
      Tingzhi Mao$^{1}$\sthanks{Tingzhi Mao has joined SCSE MICL Lab, NTU, Singapore as exchange students. This work is supported by the National Key R\&D Program of China (2017YFB1402101), Natural Science Foundation of China (61663044, 61761041), Hao Huang is correspondence author.},
      Yerbolat Khassanov$^{2,3}$,
      Van Tung Pham$^2$,
      Haihua Xu$^2$, \\
      Hao Huang$^1$, 
      Aishan Wumaier$^1$,
      Eng Siong Chng$^{2}$
    }%
}

\address{
$^1$School of Information Science and Engineering, Xinjiang University, Urumqi, China\\
$^2$School of Computer Science and Engineering, Nanyang Technological University, Singapore\\
$^3$ISSAI, Nazarbayev University, Kazakhstan
}

\begin{document}
\ninept

\maketitle

\begin{abstract}
Automatic speech recognition (ASR) for under-represented named-entity (UR-NE) is challenging due to 
such named-entities (NE) have insufficient instances and poor contextual coverage in the training data to learn reliable estimates and representations\footnote{In this paper, UR-NE refers to the named-entity (NE) words that have low-frequency count, say, the count is in [1, 9] in this work, or do not appear in the training data at all, i.e. the count is 0.}.
In this paper, we propose approaches to enriching UR-NEs to improve speech recognition performance.
Specifically, our first priority is to ensure those UR-NEs to appear in the  word lattice if there is any.
To this end, we make exemplar utterances for those UR-NEs according to their categories (e.g. location, person, organization, etc.), ending up with an improved language model (LM) that boosts the UR-NE occurrence in the word lattice.
With more UR-NEs appearing in the lattice, we then boost the recognition performance through lattice rescoring methods.
We first enrich the representations of UR-NEs in a pretrained recurrent neural network LM (RNNLM) by borrowing the embedding representations of the rich-represented NEs (RR-NEs), yielding the lattices that statistically favor the UR-NEs.
Finally, we directly boost the likelihood scores of the utterances containing UR-NEs and gain further performance improvement.
\end{abstract}

\begin{keywords}
Speech recognition, named entity recognition, word embeddings, lattice rescoring 
\end{keywords}

\section{Introduction} \label{sec:intro}
With the surge of voice-enabled applications in smart devices, the correct recognition of under-represented named-entity (UR-NE) became a paramount task, especially for the downstream applications that make use of automatic speech recognition (ASR) outputs~\cite{peyser2020improving,peyser2020improving2}.
The UR-NEs constitute the essential details of the utterance such as person, location, and organization names.

However, speech recognition for under-represented named-entity (UR-NE) words is challenging. This is because those UR-NEs
rarely occur in the training corpus, and they are also lack of contextual information, resulting in weaker language and acoustic models on such NEs.

To boost the recognition performance of those UR-NEs, our first priority is to ensure that they appear in the decoded lattice.
We then propose a series of approaches to extract the final one-best hypothesis that contains UR-NEs if there is any.

Specifically, 
we generate exemplar utterances for the UR-NEs to boost N-gram language model (LM) according to the predefined category of the NEs. We found that the UR-NE occurrence in the decoded lattice significantly improved with the help of UR-NE enriched N-gram LM. After that, we first rescore lattice by a UR-NE-enriched recurrent neural language model (RNNLM)~\cite{Khassanov2019}, and then directly boost the utterances that contain UR-NEs in lattices.

The paper is organized as follows. Section~\ref{sec:prior-work} briefs the prior related work.
In Section~\ref{sec:data-spec}, we are briefing data specification for the experiment. We then detail the proposed approaches to enriching representations for UR-NEs in Section~\ref{sec:enrich-approach}. Section~\ref{sec:exp-res} is for our experimental setup and results. 
We draw conclusions in Section~\ref{sec:con}.

\section{Related work}~\label{sec:prior-work}
In ASR community, a common practice is `` the more data, the better".
However, for those UR-NEs, even huge data cannot guarantee they are covered or sufficiently covered. Prior work is mostly focused on
out-of-vocabulary (OOV) words~\cite{he2014subword,egorova2018out,thomas2019detection}.
However, 
performance would be sub-optimal  by simply adding those OOV words to the ASR dictionary.
This is  because those OOV words have no context information, which is essential to the decode utterances containing OOV words. Another effort direction on rare word recognition is to use diversified LMs~\cite{Khassanov2019,khassanov2018unsupervised,mikolov2012subword,kim2016character,xu2018neural} to rescore lattice/N-best hypotheses, hopefully to achieve improved results. For example,~\cite{mikolov2012subword,kim2016character,xu2018neural} proposed to employ hybrid word/subword tokens as input and output units of neural LMs.
On the other hand,~\cite{Khassanov2019,khassanov2018unsupervised} proposed to augment the representations of rare words in embedding matrices of pretrained word-level neural LM.
The success of the above-mentioned methods relies on 
a presumption that those rare words do appear in the lattice which are often generated with the help of a cheaper N-gram LM. However, such assumption is not always guaranteed.

In this work, to conduct Singapore street NE recognition, we do not rely on much extra text or acoustic data. Instead, we fully exploit our training data to discover the pattern of those utterances containing NEs that are rich-represented (RR). We employ those RR-NE utterances as pool to generate exemplar utterances for those UR-NEs according to the NE category correspondence. We demonstrate such an N-gram LM boosted with the exemplar utterances can be significantly helpful on 
the UR-NE occurrence in the resulting lattice. We further apply rescoring methods on the boosted lattice to achieve improved results.

\section{Data Specification}~\label{sec:data-spec}
To train the ASR, we utilize National Speech Corpus (NSC)~\cite{Koh2019nsc} developed to advance the speech-related technologies in Singapore.
The NSC consists of three main parts: 1) read speech using phonetically balanced scripts, 2) read speech featuring words pertinent to the Singapore context, containing a lot of NEs, and 3) conversational speech.
To evaluate the proposed approaches, we use the subset of the third part (NSC-part3) as training data and a small portion of the second part (NSC-part2) as an evaluations set.
In addition, we also use SG-streets\footnote{https://github.com/khassanoff/SG\_streets} data set as an additional evaluation set.
The SG-streets data set consists of six recordings where Singaporean students read English passages containing Singapore street NEs.
The detailed data description is shown in Table ~\ref{tab:data}.
From Table~\ref{tab:data}, we can see the two test sets
contain a lot of NEs. This is particularly true for the NSC-part2 test set. It contains a lot of sentences
with NEs, for instance, utterances like ``\texttt{please look for Makaila when you reach Kallang wave mall}". 
Besides, the OOV rate related to the training data is also quite high for both test sets.
This is because the training data is conversational data, and the vocabulary is usually small~($\sim$20k in our case). 

\captionsetup[table]{labelsep=space}
\begin{table}[th]
    \renewcommand\arraystretch{1.0}
    \caption{The overall data set specification}\label{tab:data}
    \setlength{\tabcolsep}{2.5mm}
    \centering
    \begin{tabular}{l|c|c|c} 
        \toprule[1.0pt]
        \multirow{2}{*}{Category}       & \multirow{2}{*}{Train}    &\multicolumn{2}{c}{Test}   \\ \cline{3-4}
                                        &                           & NSC-part2 & SG-streets    \\ 
        \midrule
        Speakers                        & 482                       & 76        & 6             \\ 
        Duration (hrs)                  & 100                       & 1.6       & 1.0             \\
        Utterances                      & 137,058                   & 1,176     & 517           \\ 
        NE rate (\%)                    & 0.62                      & 19.33     & 7.82          \\
        OOV rate (\%)                   & -                        & 12.93     & 8.25   \\
        \bottomrule[1.0pt]
    \end{tabular}
\end{table}



\section{Approaches to Enriching UR-NE}~\label{sec:enrich-approach}
\subsection{Enrich UR-NEs with Exemplar Utterance for Language Modeling}~\label{sub:exemplar}
As mentioned, UR-NEs refer to the NEs whose count falls in, say, [0, 9], in the training data. Here, zero count means out-of-vocabulary words.
As indicated in Table~\ref{tab:data}, 
the two test sets not only contain high NE rate but also contain high OOV rate; and most of the OOV words belong to NEs.
Specifically, the NEs mainly falls in 6 categories, that is, location, person, country, company, organization, and city. Linguistically, the NEs are mostly from Mandarin, Cantonese, Hokkien, and Malay languages.

Simply adding all those UR-NEs into the ASR lexicon is a natural choice, however, 
recognition improvement won't be fully unleashed.
This is because the UR-NEs  are still lack of linguistic context that is important for LM.
To address this problem, we propose to generate exemplar utterances for them and merge such exemplar utterances to boost the language model.

In practice, we manually classify the NEs in our training data into two parts according to their count. We name those NEs with higher count (say, count in $[10, +\infty)$) as RR-NE. Simultaneously, we label the NEs with 6 categories respectively.
We then build \texttt{hash tables} for the NEs and corresponding exemplar utterances that are randomly selected from the training transcript. 

Given those UR-NEs, we label them with one of the 6 categories as mentioned above, we then  search the category by looking-up the \texttt{hash tables} we built, yielding  exemplar utterances for those RR-NEs. By simply substitute RR-NEs with the UR-NEs, we obtain corresponding exemplar utterances for the UR-NEs. Finally,
with more exemplar utterances for the UR-NEs, we update the N-gram LM to decode.

\begin{figure}[t]
\begin{tabular}{cc}
\begin{minipage}[t]{0.48\linewidth}
    \includegraphics[width = 1\linewidth]{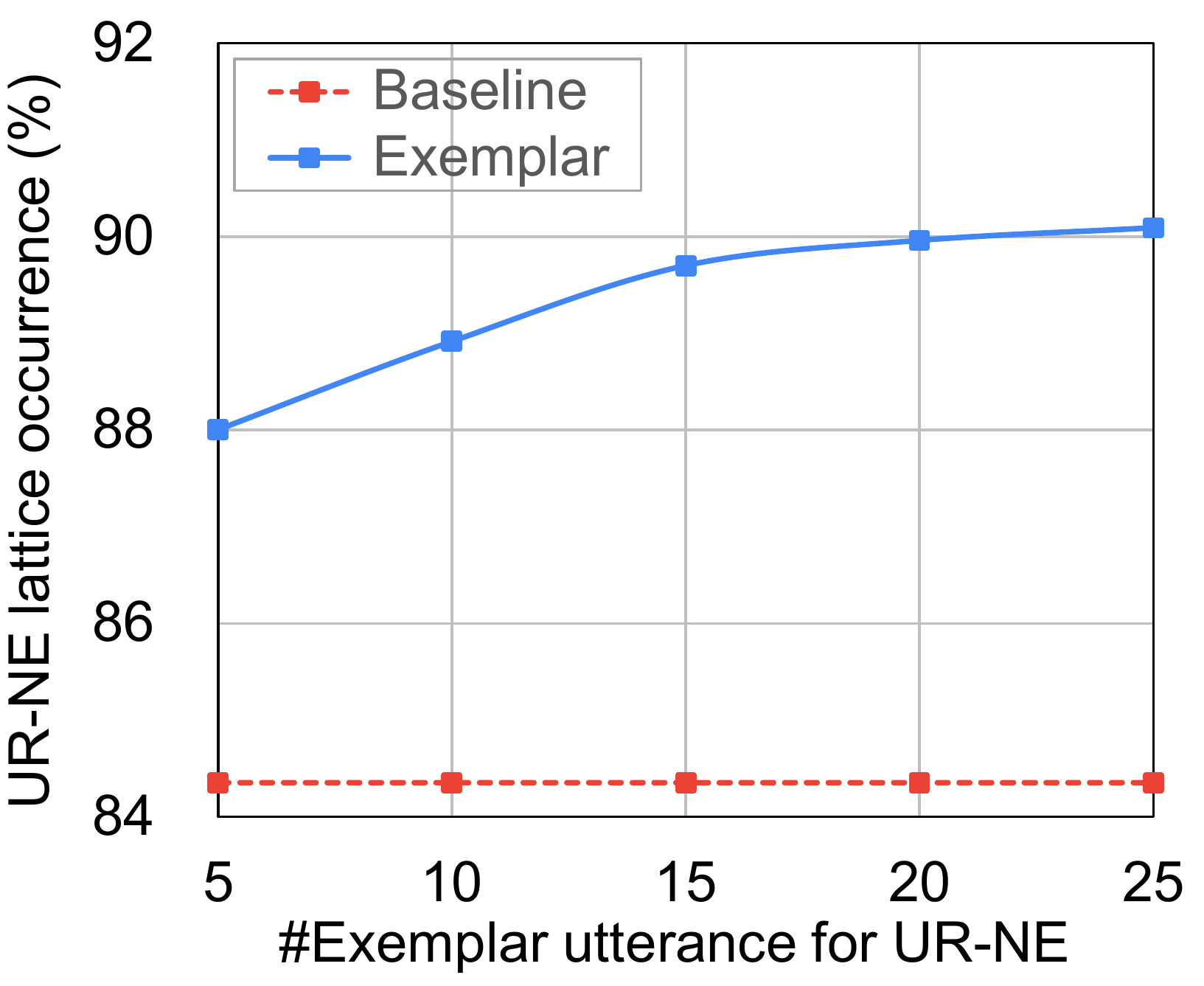}
    \centerline{(a) NSC-part2}\medskip
\end{minipage}
\begin{minipage}[t]{0.48\linewidth}
    \includegraphics[width = 1\linewidth]{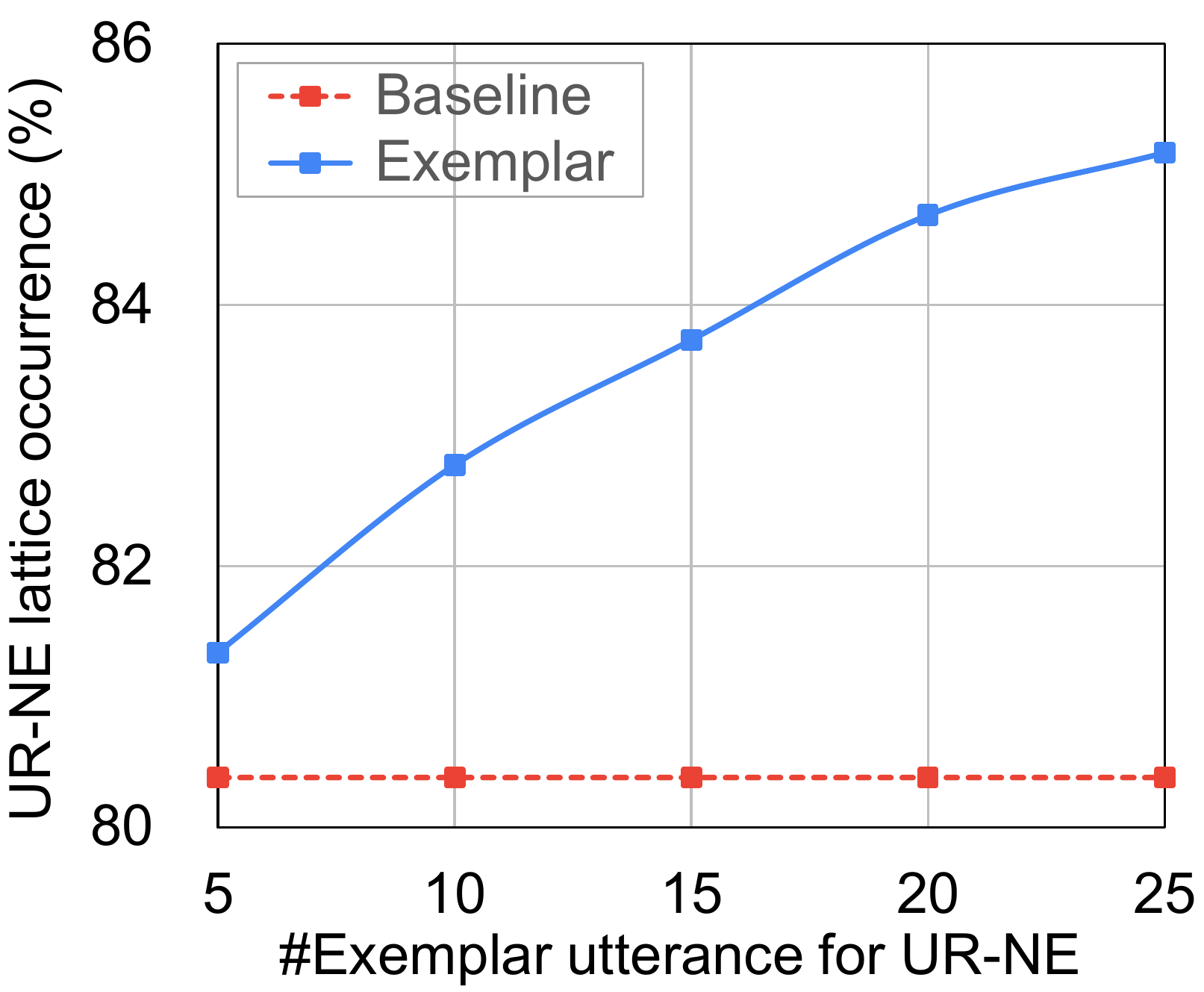}
    \centerline{(b) SG-streets}\medskip
\end{minipage}
\end{tabular}
\caption{Effectiveness of using exemplar utterances on
the UR-NE occurrence in lattices. Here, UR-NE count falls in [0, 10) in training data, and all test NEs are included in ASR lexicon for the baseline.  }\label{fig:ur-ne-lat-occ}
\end{figure}

Figure ~\ref{fig:ur-ne-lat-occ} illustrates the effectiveness of using exemplar utterance on UR-NE occurrence in the lattices for the 2 test sets in Table~\ref{tab:data}. As is clearly shown in Figure ~\ref{fig:ur-ne-lat-occ}, even with 5 exemplar utterances for UR-NE can lead to significant UR-NE occurrence in the lattices.

\begin{figure}[t]
\begin{tabular}{cc}
\begin{minipage}[t]{0.48\linewidth}
    \includegraphics[width = 1\linewidth]{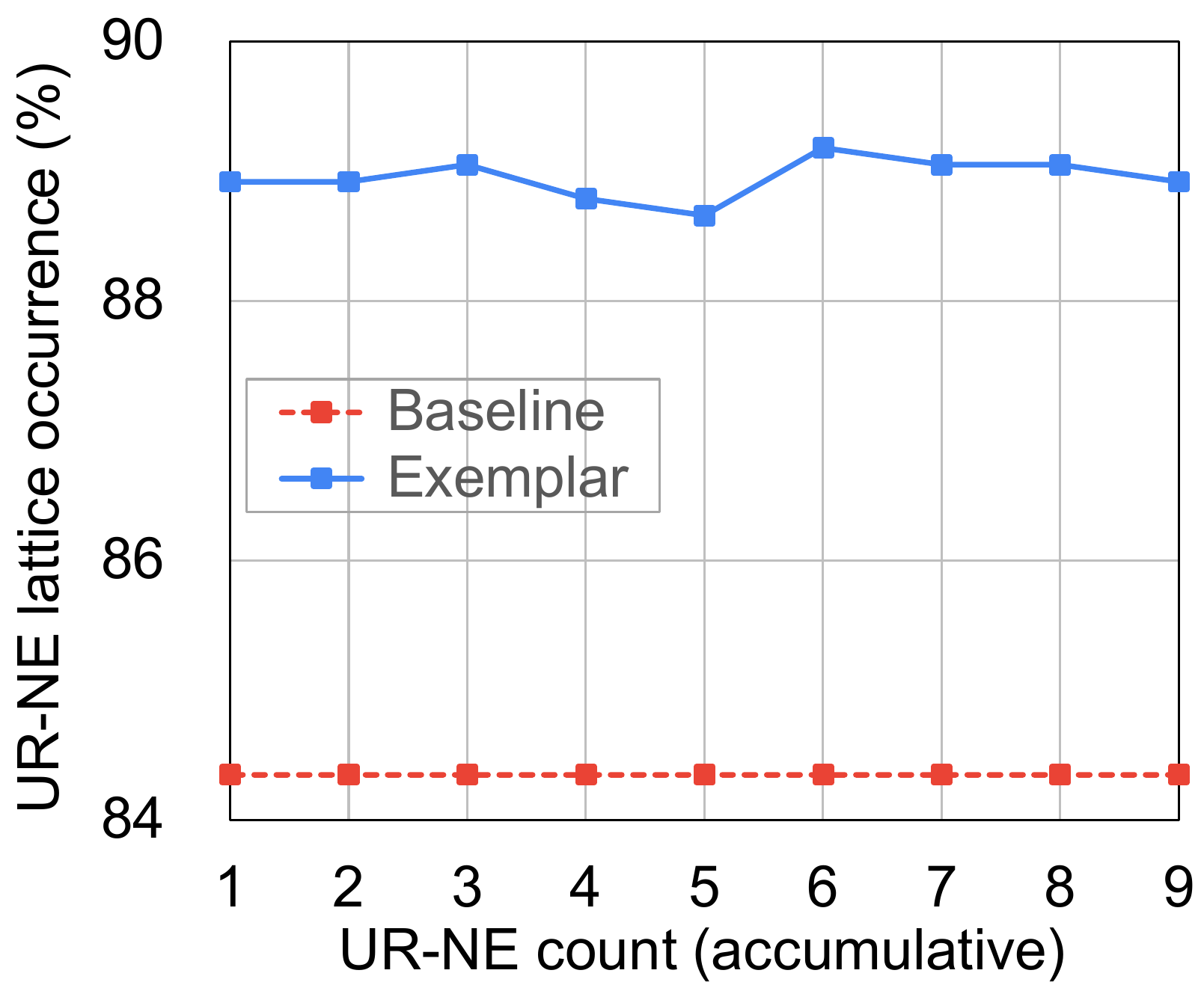}
    \centerline{(a) NSC-part2}\medskip
\end{minipage}
\begin{minipage}[t]{0.48\linewidth}
    \includegraphics[width = 1\linewidth]{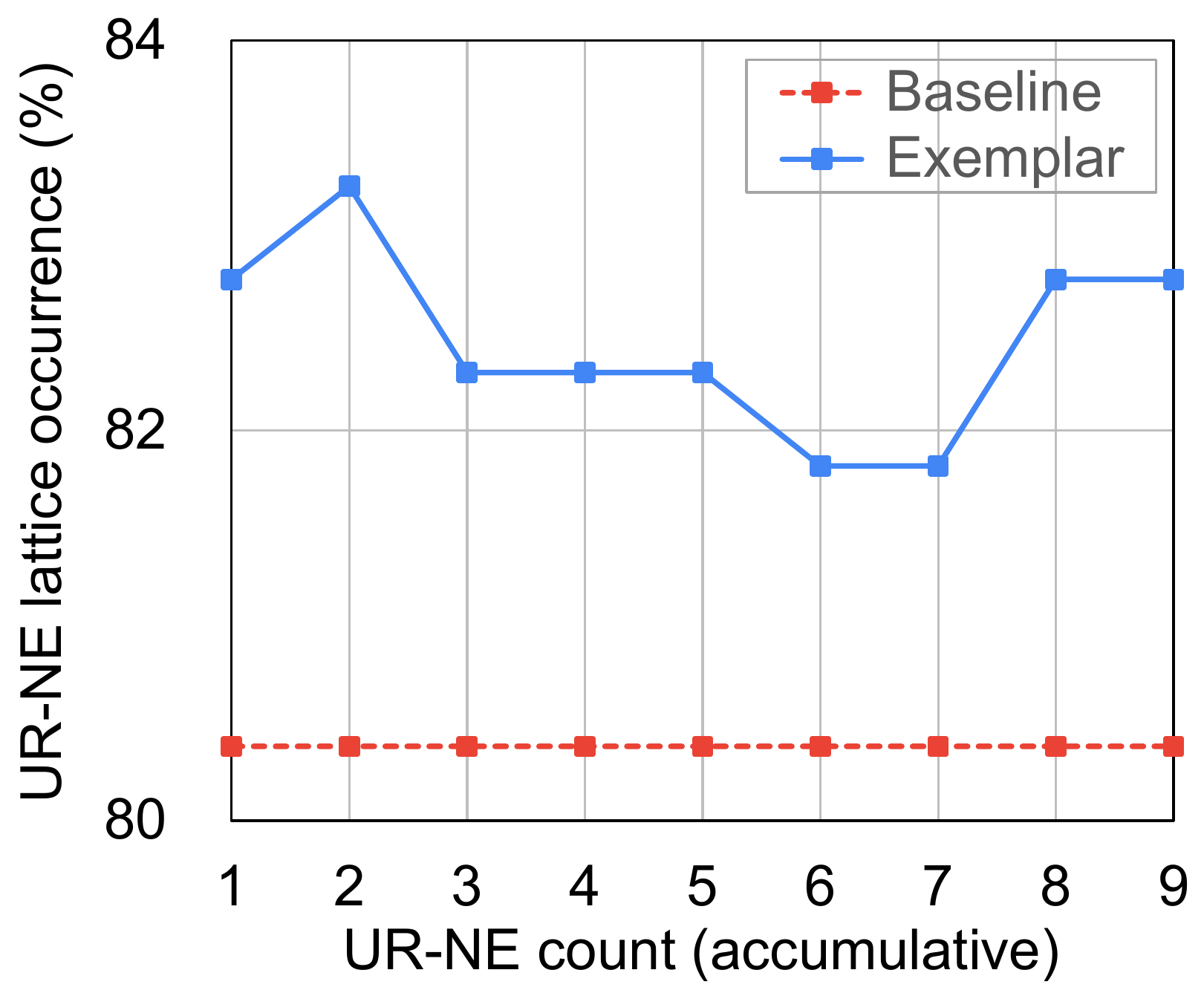}
    \centerline{(b) SG-streets}\medskip
\end{minipage}
\end{tabular}
\caption{NE occurrence in lattices versus the upperbound count of UR-NE. Here the denominator is the number of NEs whose frequency counts are in [0, 9]. }\label{fig:occ-vs-thresh}
\end{figure}

Meanwhile, we are also curious about on which part of UR-NE the exemplar method has most impact. To do this, we divide the UR-NEs into several groups according to their counts in training data. Specifically, the count of the overall NEs we are considering lies in [0, 9], we divide them into 9 groups, we then analyze the NE lattice occurrence  for each sub-group accumulatively, that is, each sub-group [0,b] means we consider the NEs with count in the corresponding level, and $b\in[0,9]$. Figure~\ref{fig:occ-vs-thresh} plots the NE lattice occurrence versus NE counts in training data.
As is observed from Figure~\ref{fig:occ-vs-thresh},
the exemplar method only has significant impact on those extremely under-represented NEs in training data, specifically, whose count lies in [0,1], namely, out-of-vocabulary and single count NEs.

\begin{figure}[h]
\begin{tabular}{cc}
\begin{minipage}[t]{0.48\linewidth}
    \includegraphics[width = 1\linewidth]{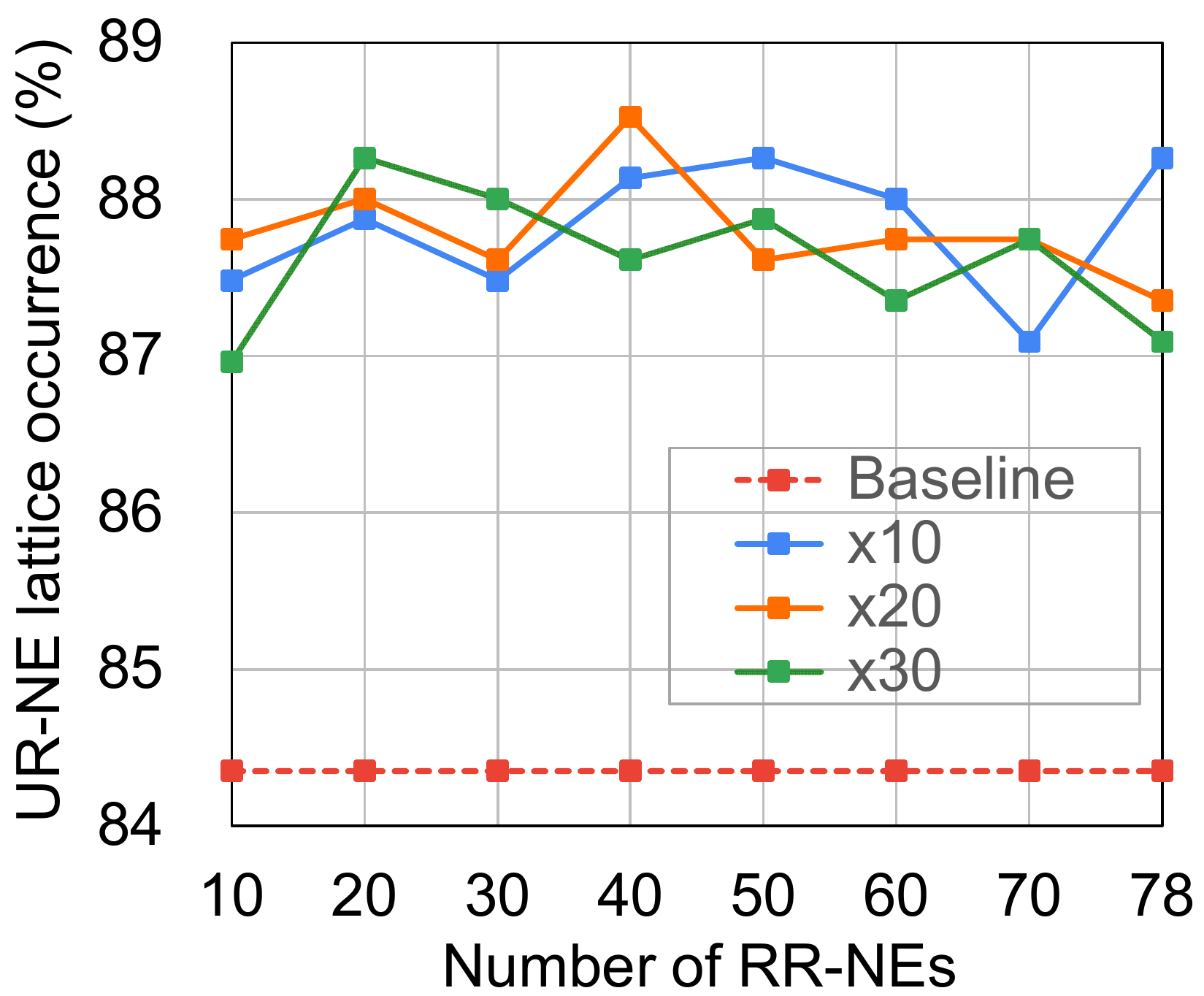}
    \centerline{(a) NSC-part2}\medskip
\end{minipage}
\begin{minipage}[t]{0.48\linewidth}
    \includegraphics[width = 1\linewidth]{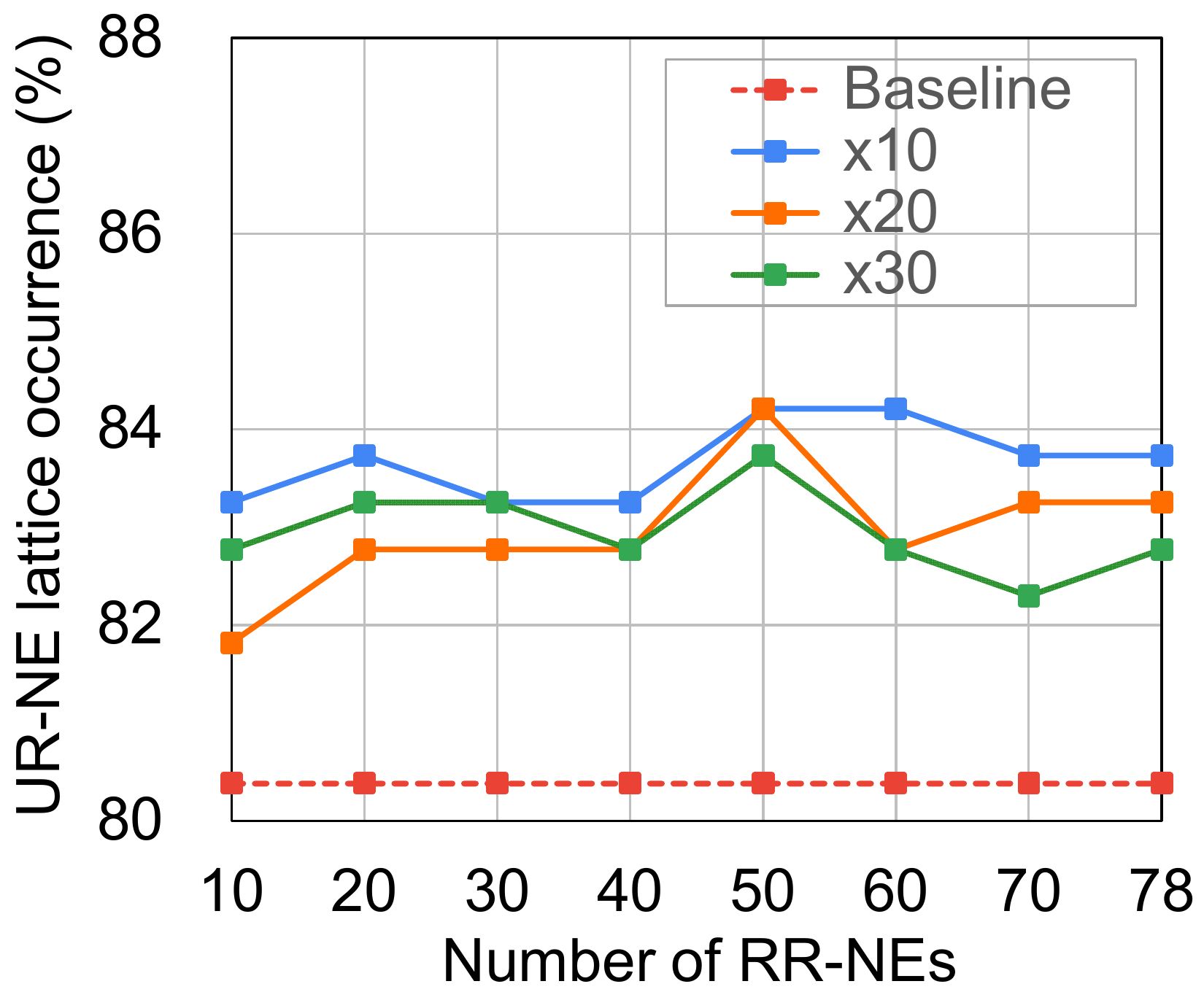}
    \centerline{(b) SG-streets}\medskip
\end{minipage}
\end{tabular}
\caption{
UR-NE lattice occurrence~(\%) versus number of RR-NEs that are employed to generate source exemplar utterances
}\label{fig:rr-ne-utt-pool}
\end{figure}

Thirdly, taking  what are suggested in both Figure~\ref{fig:ur-ne-lat-occ} and Figure~\ref{fig:occ-vs-thresh} into account, now we only
consider the UR-NEs that are either zero or single count in the training data,  and  generate 10  exemplar utterances for them. We now want to see how the NE lattice occurrence will be affected by the source utterance pool that is generated with the RR-NEs. Figure~\ref{fig:rr-ne-utt-pool}
plots the curve on the 2 test sets. In Figure~\ref{fig:rr-ne-utt-pool}, we select 10-78 RR-NEs from the training data. For each RR-NE, we collect 10, 20, and 30 source utterances respectively. From Figure~\ref{fig:rr-ne-utt-pool}, we don't observe obvious difference of effect on UR-NE lattice occurrence for each case.

Based on the above analysis on the proposed exemplar method,
we use following settings for the  experiments in Section~\ref{sec:exp-res}. We randomly select 20 RR-NEs from the training data, and for each RR-NE, we  select 30 utterances, amounting to $\sim$600 in total as utterance pool. After that, 
we only boost those UR-NEs  whose counts are in [0, 1] in training data, and for each UR-NE, we  randomly generate 10 exemplar utterances from the utterance pool.

\subsection{Enrich UR-NE for lattice rescoring}
\label{sub:ur-ne-lattice-rescore}
\subsubsection{UR-NE-enriched RNNLM lattice rescoring}~\label{subsub:rnnlm-rescore}
RNNLM lattice rescoring is an effective approach to boosting ASR performance  as a  post-processing strategy~\cite{xu2018neural,Xu2018APR}. To extract better results from lattice in Section~\ref{sub:exemplar}, we propose an improved RNNLM that enrich  the  representations for those UR-NEs following the prior work~\cite{Khassanov2019}. Specifically, 
we enrich the embedding vectors of UR-NEs using embedding vectors of 
the RR-NEs while  we keep the parameters of pre-trained RNNLM unchanged.  The formula is as follows:
\begin{equation}
  \label{eqn:enrich}
    \hat{e}_u=\frac{e_u+\sum_{e_{c}\in\mathcal{C}_r}m_{c}e_{c}}{|\mathcal{C}_r|+1}
\end{equation}
where $C_r$ is the overall embedding vector of RR-NEs set, $e_u$ is UR-NE embedding and $\hat{e}_u$ is the enriched representation of $e_u$. $m_c$ is a metric to weigh the relevance of RR-NEs to the corresponding UR-NE. In this paper, $m_c$ is 0.7 if two NEs are in the same categories, and 0.3 for the other case.

\begin{figure}[t]
\begin{tabular}{cc}
\begin{minipage}[t]{0.48\linewidth}
    \includegraphics[width = 1\linewidth]{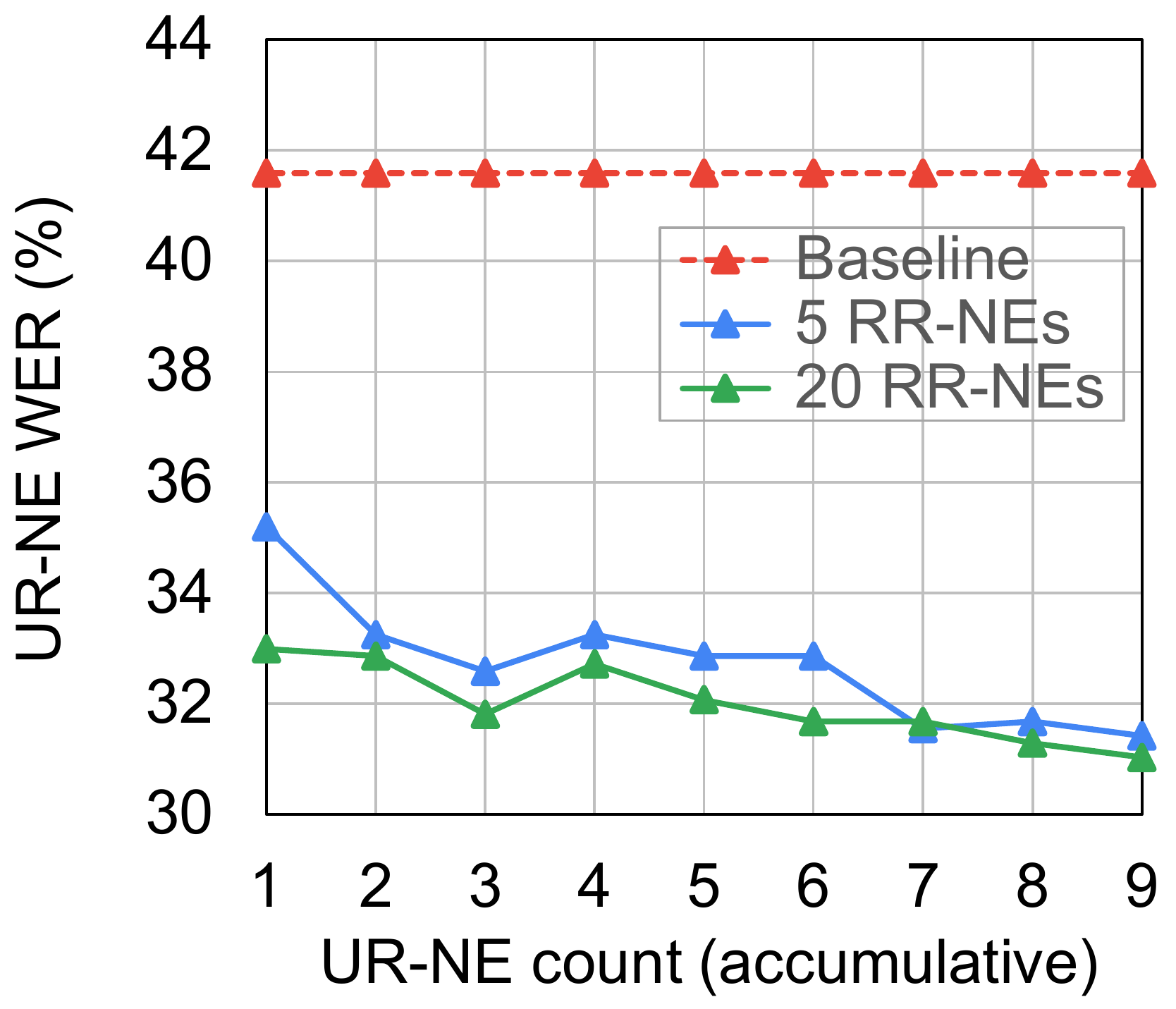}
    \centerline{(a) NSC-part2}\medskip
\end{minipage}
\begin{minipage}[t]{0.48\linewidth}
    \includegraphics[width = 1\linewidth]{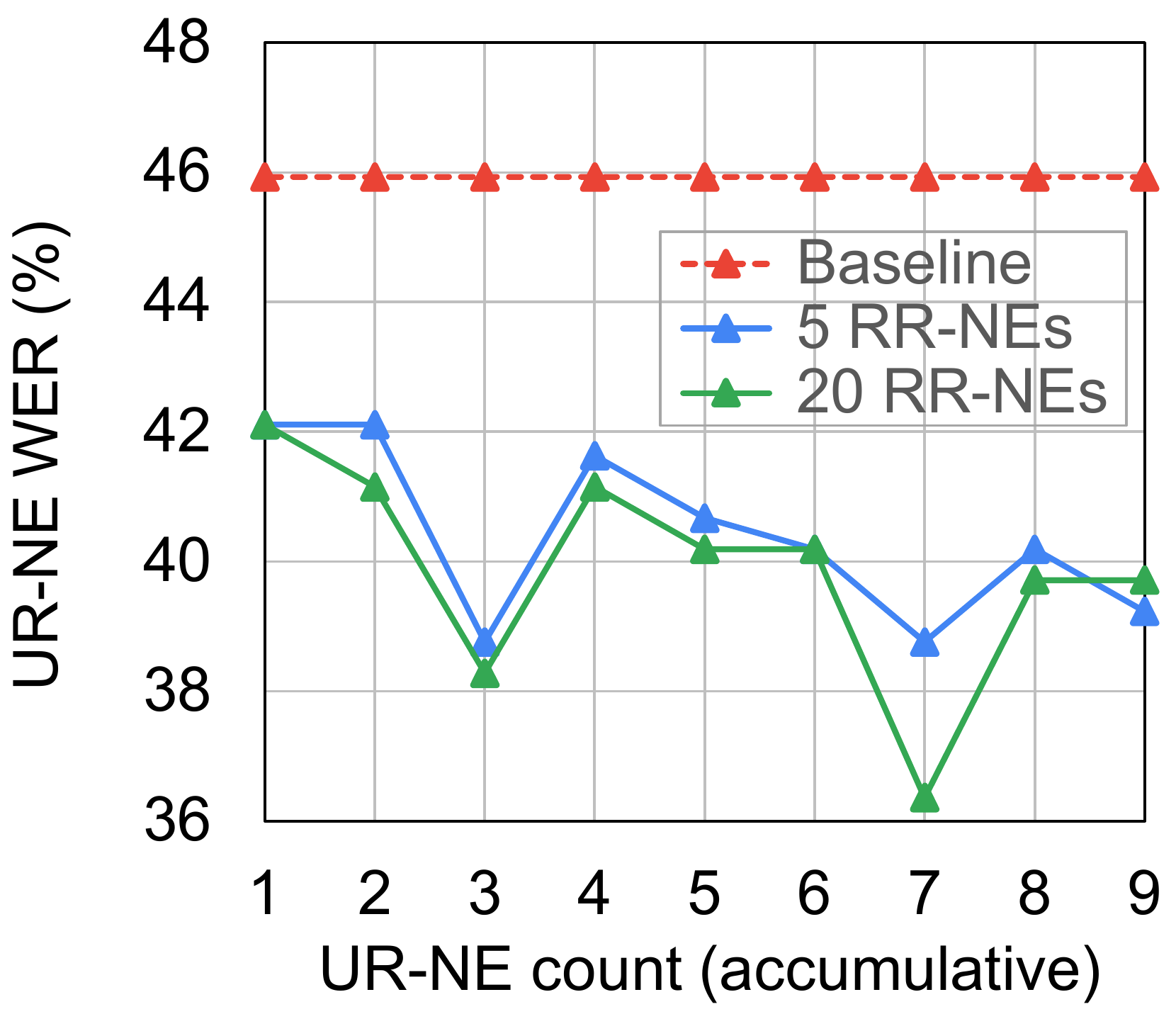}
    \centerline{(b) SG-streets}\medskip
\end{minipage}
\end{tabular}

\begin{tabular}{cc}
\begin{minipage}[t]{0.48\linewidth}
    \includegraphics[width = 1\linewidth]{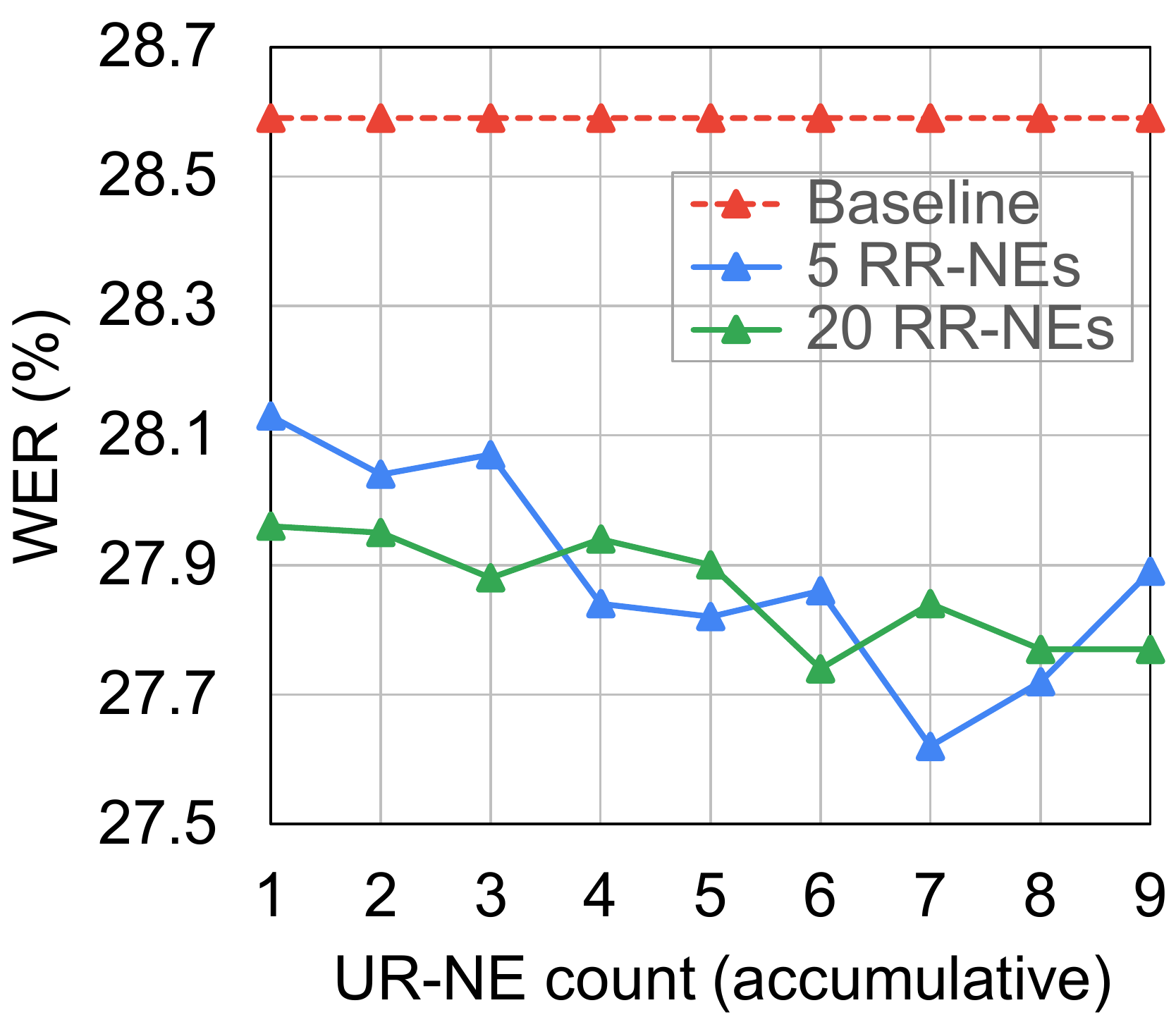}
    \centerline{(c) NSC-part2}\medskip
\end{minipage}
\begin{minipage}[t]{0.48\linewidth}
    \includegraphics[width = 1\linewidth]{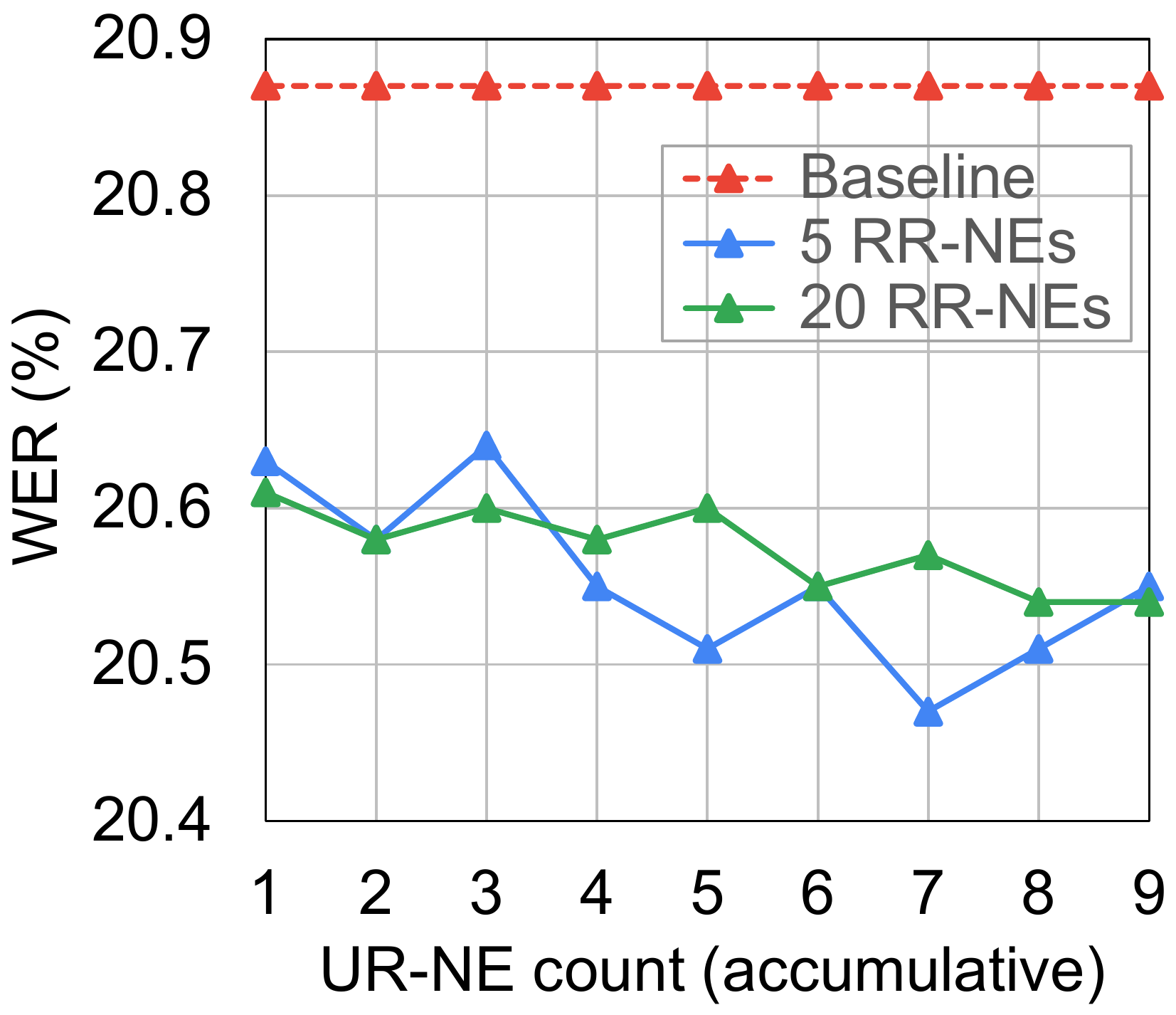}
    \centerline{(d) SG-streets}\medskip
\end{minipage}
\end{tabular}
\caption{UR-NE WER (\%) and overall WER (\%) versus different UR-NEs that are enriched with Equation~\ref{eqn:enrich}.}\label{fig:rnnlm-enriched-ne-wer}
\end{figure}

For the proposed method (for simplicity from now on, we also name it as \texttt{RNNLM-enriched} method), the success of Equation~\ref{eqn:enrich} is dependent on 2 factors, that is, how many RR-NEs and UR-NEs are involved. Besides, we also prefer the overall WER and NE-WER improvement is positively correlated.
Figure~\ref{fig:rnnlm-enriched-ne-wer} shows both  NE-WER and WER curves versus UR-NEs that are defined by their count in training data, with different RR-NE embeddings that are employed in Equation~\ref{eqn:enrich}. From Figure~\ref{fig:rnnlm-enriched-ne-wer}, we see that by using very small RR-NE embeddings (5 RR-NEs) to enrich those UR-NEs whose count is $\sim$7,  both NE-WER and WER are consistently improved.
We note that the baselines here are the corresponding normal RNNLM resoring results. More importantly, the definitions of UR-NEs and RR-NEs  here  are separated with corresponding definitions in Section~\ref{sub:exemplar}, that is, both methods can choose completely different UR-NEs and RR-NEs to enrich. In Section~\ref{sec:exp-res}, we always choose 5 RR-NE embeddings to enrich those UR-NEs whose count in [0,7].

\subsubsection{UR-NE-biased lattice rescoring}
\label{subsub:ur-ne-biased}
Even with the rescored lattice by the RNNLM-enriched method, it cannot be guaranteed the UR-NEs to appear in the 1-best hypothesis of the utterances in the end. This is because the hypothesis containing UR-NEs might have too low likelihood scores. Therefore, 
we propose a UR-NE-biased lattice rescoring method (For simplicity, we also call it as \texttt{Lattice boosting} method) to boost the hypothesis containing UR-NEs.

Practically, we are doing what follows in order. We treat all UR-NEs as keywords.
We build a trivial finite-state-transducer (FST)~\cite{Allauzen2007OpenFstAG,Riley2009OpenFstAO} for the entire UR-NE set. For a given word lattice, we first perform keyword search~\cite{latticeindex2011}  by composing the UR-NE FST with the transformed lattice. This determines if any UR-NE exists  and corresponding location in the lattice. If there is UR-NE, we 
extract the best path/hypothesis containing the UR-NE. It is simply a forward-backward path search operation on the lattice.

\begin{table*}[!ht]
\caption{The overall WERs (\%) and NE-WERs (\%) with the proposed methods. The NE-WER is computed only for the NEs whose count is in [0, 9]. Here, zero  mean the NEs are absent from the training data, while [1,9] refers to the rare case. ``Lattice boosting" means we perform UR-NE-biased lattice rescoring in Section~\ref{subsub:ur-ne-biased}. } \label{tab:all-results}
\renewcommand\arraystretch{1.05}
\setlength{\tabcolsep}{2mm}
\centering
\begin{tabular}{l|l|c|c|c|c|c|c|c|c} 
\toprule[1.0pt]
\multirow{3}{*}{ID} & \multirow{3}{*}{System}   & \multicolumn{4}{c|}{NSC-part2}                                        & \multicolumn{4}{c}{SG-streets} \\\cline{3-10}
                    &                           & \multirow{2}{*}{WER~(\%)}  & \multicolumn{3}{c|}{NE-WER (\%)}                   & \multirow{2}{*}{WER~(\%)}  & \multicolumn{3}{c}{NE-WER~(\%)} \\\cline{4-6}\cline{8-10}
                    &                           &                       & Rare          & Absent        & ALL           &                       & Rare          & Absent        & ALL \\ 
\midrule[0.75pt]
S1                  & Baseline                  & 30.14                 & 30.81         & 46.05         & 42.37         & 22.12                 & 18.57         & 58.99         & 45.45 \\ \hline
S2                  & S1 + Exemplar utterance   & 27.35                 & 28.11         & 36.43         & 34.42         & 19.85                 & 15.71         & 56.83         & 43.06 \\
S3                  & S2 + RNNLM                & 26.80                 & 28.11         & 35.22         & 33.51         & 19.41                 & 18.57         & 55.40         & 43.06 \\
S4                  & S2 + RNNLM-enriched       & \textbf{26.67}        & 26.49         & 32.65         & 31.16         & \textbf{19.35}        & 17.14         & 50.36         & 39.23 \\
S5                  & S4 + Lattice boosting     & 29.88                 & \textbf{11.89}& \textbf{20.79}& \textbf{18.64}& 23.04                 & 7.14          & \textbf{30.94}& \textbf{22.97} \\ \hline
S6                  & S1 + RNNLM                & 28.59                 & 30.27         & 45.19         & 41.59         & 20.87                 & 21.43         & 58.27         & 45.93 \\ 
S7                  & S6 + RNNLM-enriched        & 27.62                 & 24.86         & 33.68         & 31.55         & 20.47                 & 14.29         & 51.08         & 38.76 \\
S8                  & S7 + Lattice boosting     & 29.69                 & 12.97         & 23.54         & 20.99         & 22.25                 & \textbf{4.29} & 33.81         & 23.92 \\
\bottomrule[1.0pt]
\end{tabular}
\end{table*}

\section{Experimental Setup and results}~\label{sec:exp-res}

Experiments are conducted with \texttt{Kaldi} toolkit~\footnote{https://github.com/kaldi-asr/kaldi}. The acoustic models are  
the factorized time-delay neural network (TDNN-F)~\cite{Povey2018SemiOrthogonalLM}, 
trained with lattice-free maximum mutual information (LF-MMI)~\cite{Povey2016PurelySN}
criterion. All are a 6-layer convolutional neural network topped with 11-layer TDNN-F network with each layer having 1536 input neurons and 256 bottleneck output neurons respectively.
The ASR lexicon is position-dependent grapheme lexicon~\cite{Legrapheme2019}, and the vocabulary is
21.7k. Our grapheme lexicion can make comparable results with the conventional phonetic lexicon on those in-vocabulary words, while it yields better results on the NE recognition.
To realize OOV-free on either test set, we collect $\sim$3000 NEs from  website\footnote{https://geographic.org/streetview/singapore}, which covers the NE-related
OOV words on either test set.
We use 4-gram LM trained with training transcript to perform first-pass decoding to generate lattice. After that, we employ RNNLM~\cite{xu2018neural} to conduct lattice rescoring. To achieve a desired performance, we also employ both speed ~\cite{Ko2015AudioAF,Peddinti2015JHUAS}, as well as spectral augmentations~\cite{Park2019SpecAugmentAS} simultaneously.

Table~\ref{tab:all-results} presents the overall results with (S2-S5) or without (S6-S8) exemplar method in Section~\ref{sub:exemplar}  employed.
From Table~\ref{tab:all-results}, exemplar  method  achieves  better results on the final UR-NE recognition, comparing S5 with S8. On  NSC-part2, the overall NE-WERs are 18.64\% versus 20.99\%; the NE-WERs are 22.97\% versus 23.92\% on  SG-streets. 
It is particularly effective on the NEs that are absent in the training data, yielding consistent improved results on either test set. 
One point worth a notice is that it seems  the exemplar method is not perfectly coordinated with the  RNNLM-enriched method. Comparing S3 with S6, exemplar method yields obvious better results, however, after the RNNLM-enriched method is applied, the benefit is significantly reduced. As is seen from S4 versus S7, exemplar method even yields worse results (17.14\% versus 14.29\%) on 
SG-streets ``Rare" case. We conjecture it diverts the embedding ${e_u}$ in Equation~\ref{eqn:enrich}. Besides, taking a closer look into the data, we found the sentences of the SG-streets have few repetitive patterns, and they are much longer. For the NSC-part2, the patterns of the utterances containing NEs are rather restricted and repetitive, which favors for the exemplar method, since it is much easier to capture the limited context. 

Table~\ref{tab:all-results} also  shows the effectiveness of the RNNLM-enriched method on the overall WER and NE-WER improvement, with or without exemplar utterance method, as can be clearly observed in S4 and S7.  Additionally, by lattice boosting method, we can obtain remarkable NE-WER reduction on both data sets. For instance, the NE-WER is  down to 22.97\% from 39.23\% in exemplar case, and 23.92\% from 38.76\% without exemplar method on SG-streets test set. However, the overall WERs are slightly worsen, as is seen when S4 versus S5, and S7 versus S8 are respectively compared.


Finally in Table~\ref{tab:all-results} as mentioned, we observe RNNLM-enriched method  is not well coordinated with the exemplar method. We guess this is because the exemplar utterances
might divert the embedding estimate for the NR-NEs in Equation~\ref{eqn:enrich}. To verify our conjecture, we decouple the two methods, that is, we use exemplar method to generate lattice, but stick with using original training transcript to train RNNLM and let the afterwards RNNLM-enriched method not be affected by the exemplar method. Table~\ref{tab: exemplar-rnnlm-disent} reports the NE-WER results. Compared with results in Table~\ref{tab:all-results}, we notice that we make the best NE-WER results on either test sets.

\begin{table}[th]
\caption{NE-WER results of Decoupled exemplar and RNNLM-enriched methods. S9 stands for
lattice rescoring that corresponds to S3 and S6 in Table~\ref{tab:all-results}, S10 refers to
RNNLM-enriched method, corresponding to S4 and S7 in Table~\ref{tab:all-results}, S11 refers to 
lattice-boosting method, whose counterpart is S5 and S8 in Table~\ref{tab:all-results}.} \label{tab: exemplar-rnnlm-disent}
\setlength{\tabcolsep}{1.5mm}
\centering
\begin{tabular}{l|c|c|c|c|c|c}
\toprule[1.0pt]
\multirow{3}{*}{ID} & \multicolumn{6}{c}{NE-WER (\%)}                                       \\ 
\cline{2-7}
                    & \multicolumn{3}{c|}{NSC-part2} & \multicolumn{3}{c}{SG-streets}  \\ 
\cline{2-7}
                    & Rare  & Absent & ALL           & Rare & Absent & ALL             \\ 
\midrule[0.75pt]
S9                   & 27.03 & 36.60  & 34.29         & 15.71 & 52.52 & 40.19            \\
S10                  & 24.32 & 32.30  & 30.38         & 15.71 & 47.48 & 36.84           \\
S11                  & 12.43 & 20.62  & 18.64         & 4.29 & 30.94  & 22.01           \\
\bottomrule[1.0pt]
\end{tabular}
\end{table}

\section{Conclusions}~\label{sec:con}
In this paper, we proposed a  bunch of approaches to enrich under-represented name-entities, yielding better name-entity recognition performance. To realize this, our first objective is to guarantee the occurrence of the under-represented name-entity is improved in decoded lattice. Consequently we introduce an exemplar utterance generation method, yielding an improved n-gram LM that favors for the under-represented name-entities. Though the exemplar method is rather heuristic, we demonstrated its effectiveness. To achieve better results on the improved lattice, we then employed two lattice rescoring methods. One is  the RNNLM-enriched lattice rescoring method, by
enriching embedding of the under-represented name-entity with corresponding embeddings of rich-represented name-entities. We found it is very effective to boost the overall ASR  performance, that is, with or without name-entity recognition considered. Another method is we directly favor the utterance that contains under-represented name-entities from lattice. With such a method, we obtain slightly degraded ASR results, but significantly better results on the under-represented name-entity recognition.



\bibliographystyle{IEEEbib}
\clearpage
\newpage
\bibliography{main.bib}

\end{document}